\documentclass{article}
            \usepackage{graphicx}
            \usepackage{amsmath,amssymb}
\begin{document}
            \title{Principal Fiber bundle description of number scaling for scalars and vectors:
            Application to gauge theory.}
            \author{Paul Benioff,\\
            Physics Division, Argonne National
            Laboratory,\\ Argonne, IL 60439, USA \\
            e-mail:pbenioff@anl.gov}

            \maketitle

\begin{abstract}
            The purpose of this paper is to put the description of number scaling and its effects on physics and geometry on a firmer foundation, and to make it more understandable.  A main point is that two different concepts, number and number value are combined in the usual representations of number structures. This is valid as long as just one structure of each number type is being considered.  It is not valid when different structures of each number type are being considered.  Elements of  base sets of number structures, considered by themselves, have no meaning.  They acquire meaning or value as elements of a number structure. Fiber bundles over a space or space time manifold, M, are described. The fiber consists of a collection of many real or complex number structures and vector space structures. The structures are parameterized by a real or complex scaling factor, s. A vector space at a fiber level, s, has, as scalars, real or complex number structures at the same level.  Connections are described that relate scalar and vector space structures at both neighbor M locations and at neighbor scaling levels.  Scalar and vector structure valued fields are described and covariant derivatives of these fields are obtained. Two complex vector fields, each with one real and one imaginary field, appear, with one  complex field associated with positions in $M$ and the other with position dependent scaling factors.  A derivation  of the covariant derivative for  scalar and vector valued fields gives the same vector fields. The derivation shows that the complex vector field associated with scaling fiber levels is the gradient of a complex scalar field. Use of these results in gauge theory shows that the imaginary part of the vector field associated with M positions acts like the electromagnetic field.  The physical relevance of the other three fields, if any, is not known.
            \end{abstract}

\section{Introduction}

In earlier work \cite{BenIntech,BenNova,BenFB} the effect of scaling of number systems on some quantities in physics and geometry was described.  The work was based on the description of mathematical systems of different types as structures or models that satisfy a set of axioms relevant to the  type of each system \cite{Barwise,Keisler}. As presented, the work may have been difficult to understand.

The concern of this paper is to place this work on a more secure and, hopefully, easier to understand foundation. The concept of mathematical systems as structures is retained. Structures for a system of a given type consist of a base set, a set of a few base operations, a possibly empty set of a few base relations, and a set of a few constants. Examples of structures are those for the  natural numbers, $\bar{N}=\{N,+,\times, <,0,1\}$ which satisfy the axioms for arithmetic \cite{Kaye} and the real numbers, $\bar{R}=\{R,\pm,\times, (-)^{-1},0,1\}$ that satisfy the axioms for a complete ordered field \cite{real}.

This work begins with the idea that there are two separate concepts that are combined or  identified in these number  structure representations.  These are the concepts of numbers or  labels for the elements of the base sets, $N$ or $R,$ \cite{Tegmark} and the concept of the values that the numbers have in the structures.  As long as one works with just one (standard) natural number structure $\bar{N}$ and just one real number structure, $\bar{R},$ the distinction between these concepts can be ignored.  This is the case in the usual descriptions of numbers in mathematical analysis and in the use of mathematics in physics and geometry.

The distinction between these two concepts cannot be ignored when one considers the existence of many number structures of a given type that are related by scaling factors. One finds that the  base sets, $N$ and $R,$ when considered by them selves as sets, have no meaning as sets of number values.  The elements of the sets acquire meaning only as base sets of structures satisfying relevant axioms. It turns out that a given element of $N$ or $R$  can have many different values.  The particular value depends on the structure containing the base set. The value of a base set element in a number structure is defined by the properties it has.

A simple example of this is shown in the next section.  Two natural number structures are described, one is the usual one and the other is a structure for the even natural numbers. Some of the base set elements have two different values.  They have one value in the usual structure and another in the structure for the even natural numbers.

Section \ref{NS}  extends the example of natural numbers to real and complex number and to vector spaces. It also applies to integers and rational numbers, even though they are not discussed here.  Two different types of maps between structures of a given type are described, those that are the identity on the base sets and change number values, and those that preserve number values, but are scaling permutations of the base set elements. A brief description is given  of the similarity of  assignment of meaning to the base set elements and decoding of the base set elements.

The following section gives  fiber bundle descriptions of these number and vector space structures over a manifold, $M$. Here $M$ is taken to be flat as in space time or Euclidean space. Bundles for  scalars and for vector spaces with associated scalars are described. The fibers consist of either scalar or scalar and vector space structures at all possible scaling levels. The structures at different levels all have the same values.  They differ in which base set elements have a given value. A fiber at point $x$ of $M$ has a complete set of structures at all scaling levels.

Section \ref{CS} describes connections between fibers at neighboring locations and sections as fields on the bundle.  The connections are used to describe covariant derivatives of fields over $M.$ The fields described here are different from those in gauge theories in that they have two connection dimensions.  One is the usual one over $M$.  The other is the connection over the different fiber levels.  Two types of fields are discussed. One type consists of fields whose values are scalar structures  or the product of scalar and vector structures. These are fiber elements.  The other  type  consists of fields whose values are either scalars or are products of scalars and vectors. The structures containing the field values are at different fiber locations and levels in the fibers.

Section \ref{GT} describes gauge theories based on the fiber bundle setup described here. Covariant derivatives for Abelian and nonabelian theories for $2$ dimensional vector spaces are described. Four vector gauge fields are present in addition  to the three vector bosons present in the nonabelian case. The two fields, one real and one imaginary, associated with the scaling degree of freedom, appear to be gradients of scalar fields, The other two fields, also one real and one imaginary,  are associated with the locations on $M$. The imaginary field seems to have the properties of the electromagnetic field. The physical properties, if any, of the real location field and the complex scaling field are unknown at present.

A discussion section on open problems and extensions of the results obtained here ends the paper.

This work on scaling of number and vector space structures seems to be carried out in isolation as a literature search came up empty.  However  recent papers on the relativization of arithmetic and symmetries in physics \cite{Czachor} and on the mathematical universe \cite{Tegmark} does touch on some of the ideas presented here. Also the type of scaling described here seems to be different from that in conformal theories \cite{Ginsparg}. The reason is that all number values are scaled, including  those for angles and for lengths.  However the property of scaling is such that trigonometric relations are preserved under scaling.

\section{The Basics}\label{NS}
The material in this section is basic to everything that follows.  To begin one must have some idea of what mathematics is about.  Here the view is taken that mathematics consists of a large number of structures of different types with many relations between the structures \cite{Barwise,Keisler}.  Examples of structure types are the different types of numbers, natural, integers, rational, real, complex, vector spaces of different types algebras, etc.

A structure of a given type consists of a basic set, a few basic operations, none or a few basic relations, and none or a few basic constants.  A generic representation of a structure has the form
\begin{equation}\label{BS}\bar{S}=\{S,Op, Rel,K\}.\end{equation} Here $S$ is the base set, $Op$ and $Rel$ are the sets of basic operations and relations, and $K$ denotes the set of constants, if any.  An overline as in $\bar{S}$ distinguishes a structure from the base set $S$ of the structure. Structures of a given type must satisfy a set of axioms relevant to the structure type. This is in essence the   mathematical logical description of  semantic or meaningful models of formal systems as syntactic or meaningless systems with a given set of expressions as  the formal axioms \cite{Barwise,Keisler}.

Examples of structures of different types are \begin{equation}\label{BNRC}\begin{array}{c}\bar{N}= \{N,+,\times,0,1\},\mbox{ the natural numbers}\\\\\bar{R}=\{R,\pm,\times,(-)^{-1},<,0,1\}, \mbox{ the real numbers}\\\\\bar{C}=\{C,\pm,\times,(-)^{-1},^{*},0,1\}\mbox{ the complex numbers}.\end{array}\end{equation}The natural numbers satisfy the axioms of arithmetic \cite{Kaye}, and the real and complex numbers satisfy
axioms for a complete ordered field \cite{real} and an algebraically closed field of characteristic
$0$ \cite{complex}. The inverse operation is denoted by $(-)^{-1}.$ Complex conjugation has been added as a basic operation  to the complex number structure even though it is not part of the usual  representation. These structures and those for the integers and rational numbers, represent the usual structures in wide use in physics and mathematics.

 The basic point to make is that these structures and those of other types conflate or identify two distinct concepts.  One is the description of the base set elements as numbers. The other is the values the numbers have in number structures. In other words, the base set elements, when  considered by themselves, have no intrinsic meaning or values. They acquire meaning or values only as part of a structure. To emphasize this the elements of the base sets of the structures in Eq. \ref{BNRC} are referred to a numbers.  This is distinct from the  values they have in a structure. This is seen by the observation that in different structures of the same type the values of the numbers are different. Their values depend on the structure containing them.

 A simple example illustrates this point.  Consider the natural number structure $\bar{N}.$ The elements of the base set have names, "0","1","2", etc.  The quotes denote the fact that these are element names. In $\bar{N}$ the elements $"n"$ have value $n.$

 It is clear that the even numbers, "0","2","4",etc. in the base set $N$ form a base set, $N_{2},$ of a valid natural number structure $\bar{N}^{2}$. this follows from the observation that even numbers also make a good representation of the natural numbers: one can count with them and do arithmetic with them.  The structure, $\bar{N}^{2}$, for these numbers is \begin{equation}\label{BN2}
 \bar{N}^{2}=\{N_{2},+_{2},\times_{2},0_{2},1_{2}\}.\end{equation}   In this structure $"0"$ still has value $0$.  However the element $"2"$ has value $1_{2}$.  This is different from the value of $"2"$ in $\bar{N}.$  The same difference holds for any base set element $"2n"$.  This element has value $n_{2}$ in $\bar{N}_{2}$ and value $2n$ in $\bar{N}.$

 Values of base set elements of a structure are determined by the properties they have.  These are derived from the axioms. For example the value $1_{2}$ is assigned to $"2"$ in $\bar{N}^{2}$ because it satisfies the multiplicative identity axiom in that $n_{2}\times_{2}1_{2}=n_{2}$ for all number values, $n_{2}.$

 The subscript $2$ is a structure identifying label.  It identifies structure membership of a number value.  The value $1$ is a fixed meaningful value, but without a structure identifying label one has no idea which of the base set elements has value $1$.

 This description of even number structures extends to structures $\bar{N}^{n}$ for any $n.$  The base set of this structure, $N_{n},$ consists of the numbers, $"0","n","2n","3n","4n"$, etc. The values of these elements in $\bar{N}$ is given by $"0"$ has value $0_{n}$, $"n"$ has value $1_{n}$, $"2n"$ has value $2_{n},$ etc. If $"n"$ is a number in $N_{2}$, then $"n"$ has value, $(n/2)_{2}$ in $\bar{N}_{2}.$ The number, $"3n",$ has values $3_{n}$ in $\bar{N}^{n}$, $(3n/2)_{2}$ in $\bar{N}^{2},$ and value, $3n,$ in $\bar{N}$.  Note that $\bar{N}\equiv \bar{N}^{1}.$

 An interesting point is that the number, $"0",$ is the only base set element whose value, as the additive identity, is the same in all structures.  For this element, and only this element, the concepts of number and number value can be merged. Stated differently, this is the only base set number whose value is invariant under scaling.

\subsection{Scalar structures}\label{SS}

 This distinction between number value and number also applies to the other types of numbers.  Here the main interest is in real and complex numbers as they are fundamental to physics and geometry.  Let\begin{equation}\label{BSSpm}\bar{S}= \{S,\pm,\times,(-)^{-1},q,0,1\}\end{equation} be a generic structure for real and complex numbers.  If $S=R$ for the real numbers, then $q=<,$ the less than relation.  If $S=C$ for the complex numbers , then $q=(-)^{*},$ which is complex conjugation.

  For each number, $s$  in  the base set,\footnote{There are no quotes around $s$ because almost all real and complex numbers are not nameable.} $S$, define a structure, $\bar{S}^{s}$ by \begin{equation}\label{BSs}\bar{S}^{s}=\{S,\pm_{s}, \times_{s},a_{s}^{-1_{s}},q_{s},0_{s},1_{s}\}.\end{equation}Here $a_{s},0_{s},$ and $1_{s}$ are values of specific elements of $S.$ The element of $S$ whose value in $\bar{S}^{s}$ satisfies the multiplicative axiom of identity has the value $1_{s}$ in $\bar{S}^{s}.$ To be specific, the element $s$ in $S$ has the value $1_{s}$ in $\bar{S}^{s}.$

  One sees here that, as was the case for the natural numbers, the values of the numbers in the base set $S$ depend on the structure containing $S.$  The number $sa$ in $S$ has value $sa=(sa)_{1}$ in $\bar{S}.$ It has value $a_{s}$ in $\bar{S}^{s}.$

The replacement of the structure $\bar{S}^{s}$ with those for real or complex numbers is given by
\begin{equation}\label{Sups}\bar{S}^{s}= \left\{\begin{array}{l}\bar{R}^{r}=\{R,\pm_{r},\times_{r} ,a_{r}^{-1_{r}},<_{r},0_{r}, 1_{r}\}\mbox{ if $S=R$ and $s=r$}\\\\\bar{C}^{c}=\{C, \pm_{c},\times_{c}, a_{c}^{-1_{c}},(a_{c})^{*_{c}},0_{c},1_{c}\}\mbox{ if $S=C$ and $s=c$}.\end{array}\right.\end{equation} In these structures $0_{s}$ and $1_{s}$ denote the values, $0$ and $1$, in the structures $\bar{S}^{s}.$  Note that  the operations $\pm_{s},(-)^{-1_{s}},(-)^{*_{c}}$ and the relation $<_{r}$ apply to the values of numbers in the structure, $\bar{S}^{s}.$ Examples of this are the equations, $a_{s}\times_{s}b_{s}=d_{s}$ and $a_{s}+_{s}b_{s}=e_{s}.$ Also the usual representation of $\bar{S}$ as $\bar{R}$ or $\bar{C}$ is the special case here where $s=1$ as $\bar{S}^{1}=\bar{S}.$

Let $s$ and $t$ be two different scaling parameters. Let  $\bar{S}^{s}$ and $\bar{S}^{t}$  be the two  structures associated with $s$ and $t.$ The structure $\bar{S}^{t}$ is given by Eq. \ref{BSs} with the subscript $s$ replaced by $t$. These two structures differ in the associations between elements of $S$ and their values in the different structures.  An element $sa$ in $S$ has the same value, $a_{s},$ in $\bar{S}^{s}$ as the element $ta$ has, as $a_{t},$ in $\bar{S}^{t}.$ (Subscripts on numbers denote their values in structures specified by the subscript.)

It is useful  to define value maps that map elements of $S$ to their values in structures. Let $val_{s}$ be a map that maps numbers in $S$ to their values in $\bar{S}^{s}.$ One has \begin{equation}\label{vals}val_{s}(sa)=a_{s}\end{equation} for all $a$ in $S$.  Equivalently $val_{s}(a)=a_{s}/s_{s}.$ Note that $s_{s}=val_{s}(ss)$ as $val_{s}(s)=1_{s}.$

The number, $ta$  also has a value in $\bar{S}^{s}.$ The value is given by \begin{equation}\label{valsat} val_{s}(ta)=(ta/s)_{s}=t_{s}\times_{s}{s}^{-1_{s}}\times_{s}a_{s}.\end{equation}  Note that $val_{s}(ta)=val_{s}(at).$

This result can be interpreted as expressing the value, $a_{t}$ of $ta$ in $\bar{S}^{t},$ in terms of values of numbers in $\bar{S}^{s}$. In other words \begin{equation}\label{valts}val_{s,t}(a_{t})=(t/s)_{s}a_{s}. \end{equation}Here $val_{s,t}$ maps the values of numbers in $\bar{S}^{t}$ to those of $\bar{S}^{s}$.  It is related to the maps $val_{s}$ and $val_{t}$ by \begin{equation}\label{valstval}val_{s,t}(val_{t}(ta))=val_{s}(ta)\end{equation} for all numbers in $S$.

This  mapping of number values from $\bar{S}^{t}$ to those of  $\bar{S}^{s}$  can be extended to other components of the structures.  The resulting structure is given by \begin{equation} \label{BSst}\bar{S}^{t}_{s}=\{S,\pm_{s}, \frac{s\times_{s}}{t},\frac{t(a_{s})}{s}^{-1_{s}}, \frac{tq_{s}}{s},0, \frac{t1_{s}}{s}\}.\end{equation}This equation gives the values of the components of $\bar{S}^{t}$ in terms of those of $\bar{S}^{s}$.  It is a relativization of the components of $\bar{S}^{t}$ to those of $\bar{S}^{s}.$
The scaling of the components in this structure is done to satisfy the requirement that this relativized structure satisfy the axioms for real or complex numbers if and only if the structures, $\bar{S}^{s}$ and $\bar{S}^{t},$ satisfy the axioms.

The relativization  of  $\bar{S}^{t}$ to $\bar{S}^{s}$  can be defined by a map, $Z_{S,s,t},$ from the components of $\bar{S}^{t}$ to those of $\bar{S}^{t}_{s}.$  It is a number preserving but not value preserving map.  One has \begin{equation}\label{ZSstBS}Z_{S,s,t}\bar{S}^{t}=\bar{S}^{t}_{s}
\end{equation} where \begin{equation}\label{ZSst}\begin{array}{l} Z_{S,s,t}(a)=a \mbox{ for all numbers, $a$ in $S$,}\\\\ Z_{S,s,t}(a_{t})=val_{s,t}(a_{t})=\frac{\mbox{$ta_{s}$}}{\mbox{$s$}},\;\;\; Z_{S,s,t}(\pm_{t})=\pm_{s}\\\\ Z_{S,s,t}(\times_{t})= \frac{\mbox{$s\times_{s}$}}{\mbox{$t$}}, \;\;\;Z_{S,s,t}((a_{t})^{-1_{t}})=\frac{\mbox{$t(a_{s}^{-1_{s}})$}}{\mbox{$s$}} \\\\ Z_{t,s}(q_{t})= \frac{\mbox{$tq_{s}$}}{\mbox{$s$}},\;\;\;Z_{S,s,t}(0_{t})=0_{s}=0,\;\;\; Z_{S,s,t}(1_{t})=val_{s,t}(1_{t})= \frac{\mbox{$t1_{s}$}}{\mbox{$s$}}.\end{array}\end{equation} The superscript, $t$ in $\bar{S}^{t}_{s}$ refers to the initial structure.  The subscript, $s,$ denotes the structure whose components are used, with appropriate scaling factors, to represent the components of $\bar{S}^{t}.$

The term, $tq_{t}/s,$ is  a stand in for quantities that are different in real  and complex number structures. For real number structures $tq_{t}/s$ is a stand in for $<_{t}=<_{s}.$ For complex numbers it is a stand in for $(t/s)^{*_{s}}a_{s}^{*_{s}}.$

The number preserving maps are transitive and have inverses.  One has
\begin{equation}\label{ZSstZStu}\begin{array}{c}Z_{S,s,t}Z_{S,t,u}\bar{S}^{u}=Z_{S,s,t}\bar{S}^{u}_{t}= \bar{S}^{u}_{s}=Z_{S,s,u}\bar{S}^{u}\mbox{ and} \\\\Z_{S,s,t}Z_{S,t,s}\bar{S}^{s}=Z_{S,s,t}\bar{S}^{t}_{s} =\bar{S}^{s}.\end{array}\end{equation}  Note that $\bar{S}^{s}=\bar{S}^{s}_{s}$.  $\bar{S}^{s}$ does \emph{not} equal $\bar{S}^{s}_{1}.$

To  better understand the action of $Z_{S,s,t}$ and the components of $\bar{S}^{s}_{t}$ in Eq. \ref{BSst}, set $S=C$ and $s=1$.  Let $t=c$, a complex number. The components of the structures, $\bar{C}^{c}$ and $\bar{C}_{1}^{c}$ are obtained from Eqs. \ref{Sups} and \ref{BSst}.  They are,\begin{equation}\label{BCsc}\bar{C}^{c}=\{C, \pm_{c}, \times_{c}, a_{c}^{-1_{c}},(a_{c})^{*_{c}},0,1_{c}\}\end{equation} and \begin{equation}\label{BCc} \bar{C}^{c}_{1}= \{C,\pm,\frac{\times}{c},c(a^{-1}),c(a^{*}), 0,c\}=\{C,\pm_{1},\frac{\times_{1}} {c_{1}},c_{1}(a_{1}^{-1_{1}}),c_{1}(a_{1}^{*_{1}}),0,c_{1}\}.\end{equation}

The subscript, $1,$ on the components of the righthand term of Eq. \ref{BCc} emphasizes that the unsubscripted quantities in $\bar{C}^{c}_{1}$ are values and not base set elements. It also shows that the scaling factor is $1$ for the usual complex number structure.

Scaling  by a complex number may seem strange in that a base set number, $c,$ that has a complex value  in the usual unscaled structure, $\bar{C}^{1}=\bar{C}$ has a value that is the identity in the structure, $\bar{C}^{c}_{1}.$   This follows from the scaling of the multiplication operation in $\bar{C}^{c}_{1}.$  It also follows from  Eqs. \ref{vals}-\ref{valstval}.  One has $$1_{c}=val_{c}(c) \leftrightarrow val_{1,c}(1_{c})=val_{1}(c)=(\frac{c}{1})_{1}=c_{1}.$$ This shows that $c_{1}$ is the same value in $\bar{C}^{c}_{1}$ as $1_{c}$ is in $\bar{C}^{c}.$  However $c$ is not in general real in $\bar{C}^{c}_{1}.$  This follows from Eq. \ref{ZSst} in that $t$ and $s$ can be arbitrary complex number parameters.

It is important to keep in mind that scale changing maps between scalar structures and basic operations do not commute.  One example concerns complex conjugation.  Let $a_{c}$  be a number value  in $\bar{C}^{c}.$ One can  first complex conjugate $a_{c}$ to $a_{c}^{*_{c}}$ in $\bar{C}^{c}$ and then use $val_{1,c}(a_{c}^{*_{c}})$  to map $a_{c}^{*_{c}}$ to $c(a^{*})$ in $\bar{C}^{c}_{1}.$ Alternatively one can first map $a_{c}$ to $val_{1,c}(a_{c})=c_{1}\times_{1}a_{1}$ and then take the complex conjugation.  The result is $c_{1}^{*}a_{1}^{*}.$  if $c$ is complex, $c_{1}(a_{1}^{*})\neq c_{1}^{*}a_{1}^{*}.$

Another important example concerns multiplication.  Let $a_{c}$ and $b_{c}$ be two number values in $\bar{C}^{c}.$  One can either multiply them first and then scale the result  to $\bar{C}^{c}_{1},$ or one can first scale $a_{c}$ and $b_{c}$ separately to $\bar{C}^{c}_{1}$ and then multiply.  The first option gives $$val_{1,c}(a_{c}\times_{c}b_{c})=(ca)_{1}\frac{\times_{1}}{c_{1}}(cb)_{1}=c_{1}\times_{1}(a\times b)_{1}.$$ The second option gives $$val_{1,c}(a_{c})\times_{1} val_{1,c}(b_{c})=(ca)_{1}\times_{1}(cb)_{1} =c^{2}_{1}(a\times b)_{1}.$$ The two options differ by a factor of $c$.

These two examples show that, in a given situation, one must be clear about which comes first, scaling or operation.  The big advantage of operations before scaling is that equations are preserved under scaling. For example, if $a_{c}\times_{c}b_{c}=d_{c}$, then $$val_{1,c}(a_{c}\times_{c}b_{c})=(ca)_{1}\frac{\times_{1}} {c_{1}}(cb)_{1}=c_{1}(a\times b)_{1}=val_{1,c}(d_{c})=(cd)_{1}\rightarrow a_{1}\times_{1}b_{1}=d_{1}.$$ Preservation of equations is essential if axiom validity is to be preserved under scaling.

One can also define number changing but value preserving maps, $W_{S,s,t}.$  One has \begin{equation}\label{WSst} W_{S,s,t}\bar{S}^{t}=\bar{S}^{s}\end{equation}where \begin{equation}\label{WSst1}\begin{array}{c}W_{S,s,t}(ta)= sa \mbox{ for all $a$ in $S$},\\\\  W_{S,s,t}Op_{t}=Op_{s},\;\;\; W_{S,s,t}Rel_{t}=Rel_{s},\;\;\;W_{S,s,t}K_{t}=K_{s}.\end{array}
\end{equation} As was the case for the $Z_{S,s,t}$ maps, the $W_{S,s,t}$ maps are transitive and invertible.

Structures can be combined provided they have the same subscripted components.  For example, two structures $\bar{S}^{t}_{s}$ and $\bar{S}^{u}_{s},$ Eq, \ref{BSst}, can be added or subtracted.  The resultant structures are \begin{equation}\label{BStpmu}\bar{S}^{t}_{s}\pm \bar{S}^{u}_{s}=\bar{S}^{t\pm u}_{s}.\end{equation} Subtraction with $t=u$ gives the empty structure. Structures, $\bar{S}^{t}_{s}$ and $\bar{S}^{t}_{u}$ with $u\neq s$ cannot be combined.  Structures can also be multiplied by numbers.  For example, multiplication of $\bar{S}^{t}_{s}$ by the number, $w,$ gives \begin{equation}\label{BSstw}w\bar{S}^{t}_{s}= \bar{S}^{t\times w}_{s}=\bar{S}^{tw}_{s}.\end{equation} These results extend to combinations and number multiplications of structures, $\bar{S}^{t},$ if one notes that $\bar{S}^{t}\equiv \bar{S}^{t}_{t}.$ These results will be used later on.

\subsection{Vector space structures}
As noted in the introduction, scaling of number structures extends to types of mathematical systems that include scalars in their description.  Vector spaces are an example since they are closed under scalar vector multiplication. A normed vector space based on the  scalars in $\bar{S},$ has the structure representation,
\begin{equation}\label{BV}\bar{V}=\{V,\pm,\cdot,|v|,v\}.\end{equation} Here  $\cdot$ denotes scalar vector multiplication, $|v|$ denotes the real valued norm, in $\bar{S},$ of an arbitrary vector, $v,$ in the base set, $V$.

The vector space value structure that corresponds to $\bar{S}^{s}$ is given by\begin{equation}\label{BVss}\bar{V}^{s} =\{V,\pm_{s},\cdot_{s},|v_{s}|_{s} ,v_{s}\}\end{equation} The scalars associated with $\bar{V}^{s}$ are those in $\bar{S}^{s}.$ In $\bar{V}^{s},$ $v_{s}$ is the value of the vector $sv$ in the base set $V.$ Note that the structure $\bar{V}^{1}\equiv\bar{V}$ of Eq. \ref{BV}.

The value preserving and base set preserving maps, $W_{S,s,t}$ and $Z_{S,s,t}$ of Eqs. \ref{WSst} and \ref{ZSst}, can be extended to apply to the vector spaces.  These maps are denoted by $W_{V,s,t}$ and $Z_{V,s,t}.$

The action of $W_{V,s,t}$ on the base set, $V$ of vectors is defined as follows.  For each vector $tv$  in $V,$ \begin{equation}\label{WVst}W_{V,s,t}(tv)=sv.\end{equation} The map is such that the value, $v_{t},$ of $tv$ in $\bar{V}^{t}$ is the same as is the value, $v_{s},$ of $sv$ in $\bar{V}^{s}.$ As was the case for the scalar base set preserving maps, the $W_{V,s,t}$ maps are transitive and have inverses.

A product of the maps $W_{S,s,t}$ and $W_{V,s,t}$ is defined by\begin{equation}\label{WSVst}W_{SV,s,t} \bar{S}^{t} \times\bar{V}^{t}=W_{S,s,t}\bar{S}^{t}\times W_{V,s,t}\bar{V}^{t}=\bar{S}^{s}\times\bar{V}^{s}. \end{equation} This definition accounts for the fact that the parameters, $s$ and $t$ in the structure superscripts must be the same for both vectors and scalars.  This satisfies the requirement that for each $s,$ the scalar field for $\bar{V}^{s}$ is $\bar{S}^{s}.$

 Unlike the $W_{V,s,t}$ maps, the $Z_{V,s,t}$ maps are base set vector preserving.  For each vector, $v$ in $V$, $Z_{V,s,t}(v)=v$ is the same  base set vector in both $Z_{V,s,t}V^{t}$ and $V^{s}.$  The action of $Z_{V,s,t}$ on $\bar{V}^{t}$ is defined by \begin{equation}\label{ZVstBVt}Z_{V,s,t}\bar{V}^{t}= \bar{V}^{t}_{s}=\{V,\pm_{s}, \frac{s}{t}\cdot_{s},\frac{t}{s}|v_{s}|_{s},\frac{t}{s}v_{s}\}.\end{equation}

This equation shows that a vector, $tv,$ with value $v_{t}$ in $\bar{V}^{t}$, has value $(t/s)W_{V,s,t}v_{t} =(t/s)v_{s}$ in $\bar{V}^{s}.$ The norm, $|v_{t}|_{t},$ of a vector value in $\bar{V}^{t}$ is a real number value in $\bar{S}^{t}.$  In $\bar{S}^{s}$ this number value is $$(t/s)|W_{V,s,t}(v_{t})|_{s}=(t/s)W_{S,,s,t}|v_{s}|_{t}=(t/s)|v_{s}|_{s}.$$  The representation of the scalar vector product operator value $\cdot_{t}$ in $\bar{V}^{s}$ is the same as the representation of scalar multiplication in $\bar{S}^{t}$ relativised to $\bar{S}^{s}.$

As was the case for the value preserving maps, Eq. \ref{WSVst}, one can define a product scalar vector base set preserving map by \begin{equation}\label{ZSVst}Z_{SV,s,t}(\bar{S}^{t}\times\bar{V}^{t})=Z_{S,s,t} \bar{S}^{t} \times Z_{v,s,t}\bar{V}^{t}=\bar{S}^{t}_{s}\times\bar{V}^{t}_{s}.\end{equation}This map will be of use later on.

\subsection{Relativization invariance}\label{RI}

The value structures $\bar{S}^{t}$ and $\bar{V}^{t}$ for the different values of $t$ are all equivalent.  One pair of structures is just as good as another to use in mathematics.  However one might single out the structure with $t=1$ as special and use it.  One might try to justify this choice in that it corresponds to the one in general use as $\bar{S}$ and $\bar{V}.$  Also this choice is the only one for which the concepts of number and number value coincide.

The problem with this is that one can scale the parameters, $s$ and $t$ by multiplying by some common factor, $\alpha$.  The result of this  is that $\bar{S}^{t}$ and $\bar{V}^{t}$ become $\bar{S}^{t\alpha}$ and $\bar{V}^{t\alpha}.$ In particular, $\bar{S}$ and $\bar{V}$ become $\bar{S}^{\alpha}$ and $\bar{V}^{\alpha},$ and $\bar{S}^{\alpha^{-1}}$ and $\bar{V}^{\alpha^{-1}}$ become $\bar{S}$ and $\bar{V}.$

This shows that no one of the pair of scalar and vector value structures can be singled out as the pair to use. Structures scaled by any value of $t$ can always be rescaled to $\bar{S}$ and $\bar{V}$ by appropriate choice of the rescaling factor.

However relative scaling between scaled structures is invariant under scaling the parameters, $s,t.$  This is seen by use of Eqs. \ref{BSst} and \ref{ZVstBVt} to show that \begin{equation}\label{BSValt} \begin{array}{l}\bar{S}^{\alpha t}_{\alpha s} =\bar{S}^{t}_{s} \mbox{ and } \\\\\bar{V}^{\alpha t}_{\alpha s} =\bar{V}^{t}_{s}.\end{array}\end{equation}

\subsection{Decoding }\label{DC}
The descriptions of the base sets, $S$ and $V$ and their relations to value structures, have much in common with the syntactic semantic separation of structures as described in mathematical logic \cite{Barwise}. Syntactic structures consist of meaningless strings of symbols constructed inductively from a small set of base symbol strings. A set of symbol strings is singled out as the syntactic axioms for the structure.  The set singled out depends on the type of system being considered.  Axioms are combined with the logical rules of deduction to inductively generate a sequence of symbol strings.  Each string in the sequence is a theorem, and the string generating it is a proof of the theorem.

The semantic representation  of a mathematical system is often treated as a model of the syntactic structure. The expressions are all meaningful, and elements have values or meaning.  The axioms and theorems as meaningful. representations of the syntactic expressions are all true.  Also no false statement is a  theorem.

These semantic structures are the stuff of mathematics as usually done. For example $\bar{R},$ $\bar{C}$, and $\bar{V}$, given in Eqs. \ref{BNRC} and \ref{BV} denote the real and complex number, and vector space structures.  Their properties are described in texts on mathematical analysis.  The elements of the base sets of these structures, which are taken here to be meaningless, can be identified with the values they have in these structures. This corresponds to a single decoding  or interpretation or valuation of the syntactic expressions and of the base set elements as numbers.

Here this setup is changed.  There are an infinite number of decodings or valuations of the syntactic expressions and base set elements. For the real numbers the decodings are parameterized by real numbers.  For each $r$, $\bar{R}^{r}$  is a valuation or decoding of the syntactic expressions and of the base set elements. If $r$ and $t$ are two different real numbers, then $\bar{R}^{r}$ and $\bar{R}^{t}$ are two different decodings or valuations of the base set or of syntactic expressions. The maps $W_{R,r,t}$ and $Z_{R,r,t},$ Eqs. \ref{WSst} and \ref{ZSst}, show explicitly how the two value structures are related.

Similar descriptions hold for the complex numbers.  Decodings or valuations of symbol strings and base set elements are parameterized by complex numbers.  If $c$ and $d$ are two complex numbers, then $\bar{C}^{c}$ and $\bar{C}^{d}$ are two decodings.  The maps, $W_{C,c,d}$ and $Z_{C,c,d}$ show how these structures are related.

The description for vector spaces is similar. If the associated scalar field  is real, then the decodings  or valuations are parameterized by real numbers.  If the scalar field is complex, then the decodings are parameterized by complex numbers.

\section{Fiber bundles}\label{FB}
Fiber bundles over a manifold provide a very useful framework for the description of gauge theories and other areas of physics and geometry \cite{Daniel,Pfeifer}. They can also be  adapted to include the effects of number scaling.  In general a fiber bundle  consists of a triple, \cite{Husemoller} $E,\pi,M.$ Here $E$ is the total space, $M$ is a base space, and $\pi$ is a projection of $E$ onto $M$. The inverse map, $\pi^{-1}(x),$ is the fiber, $F_{x}$ at point $x$ of $M$. In this work $M$ is taken to be a flat space such as  Minkowski space time or  Euclidean space.

The flatness of $M$ allows one to  restrict the type of bundle  considered here to  product bundles.    Vector and scalar product bundles have the form, $\mathcal{V}=M\times\bar{V},\pi_{v},M$ and $\mathcal{S}=M\times\bar{S},\pi_{s},M.$  Here $\bar{S}$ with $S=R$ or $S=C$ denotes the scalar structure for real or complex numbers.  The fiber product \cite{Husemoller} of these two bundles is given by\begin{equation}\label{MCSV}\mathcal{SV}=\mathcal{S} \oplus\mathcal{V}=M\times(\bar{S}\times\bar{V}),\pi_{sv},M.\end{equation}   The projection, $\pi_{sv}$ is related to $\pi_{s}$ and $\pi_{v}$ by \begin{equation}\label{pisvy}\pi_{sv}(y,\bar{S}\times\bar{V})= \pi_{s}(y,\bar{S})\times \pi_{v}(y,\bar{V}).\end{equation} This holds for each $y$ in $M$.

The fiber at point  $x$ is defined by\begin{equation}\label{pim1x}\pi^{-1}_{sv}(x)=x\times\bar{S} \times\bar{V}=\bar{S}_{x}\times\bar{V}_{x}.\end{equation}The right hand term emphasizes the fact that that the scalar and vector space structures are associated with a point, $x$ of $M$.  In this sense they can be said to be localized at $x.$

It is important to note that here, and in the following, fibers consist of one or more  \emph{structures} of different mathematical types at points of $M.$ They do \emph{not consist} of  just the base set of a structure. This is indicated by the bars over $S$ and $V$ in Eq. \ref{MCSV}.

The exact description of the vector space $\bar{V}_{x}$ is left unspecified. It can be based on real scalars, as in tangent spaces, or on complex scalars, as in gauge theories.  Neither the number of dimensions nor the type of vector space is chosen.  As a normed vector space it can be  an inner product  space, or even an algebra. The choice of a normed space is not even necessary.  It is done here to have a specific structure type to work with that is also of use in physics and geometry.

 The bundle $\mathcal{SV}$ can be expanded to include  scaling.   The method used here  defines, for each  scaling factor, $s,$  the bundle, \begin{equation}\label{MCSVs}\mathcal{SV}^{s}=M\times \bar{S}^{s}\times \bar{V}^{s}, p_{s},M.\end{equation} Here $\bar{S}^{s}$ and $\bar{V}^{s}$ are defined by Eqs. \ref{BSs} and \ref{BVss}.

 The bundles for different values of $s$ can be collected together into one fiber bundle, $\mathfrak{SV},$ defined by \begin{equation}\label{MFSV} \mathfrak{SV}=M\times \bigcup_{s}(\bar{S}^{s} \times \bar{V}^{s}),P,M.\end{equation}This  bundle  includes, as subbundles \cite{Husemoller}, all the $\mathcal{SV}^{s}.$ The union is over all scaling parameters, $s.$

 The fiber at point $x$ of $M$ is given by\begin{equation}\label{Pm1x}P^{-1}(x)=x\times\bigcup_{s}(\bar{S}^{s} \times \bar{V}^{s})=\bigcup_{s}(\bar{S}^{s}_{x} \times \bar{V}^{s}_{x})=\bigcup_{s}p^{-1}_{s}(x).\end{equation} This shows that the scaled structures, $\bar{S}^{s}\times\bar{V}^{s}$, in the fiber at $x$, can be regarded as local at $x.$

The description of the bundle, $\mathfrak{SV},$ has the advantage that it automatically takes care of the requirement that the scaling factor, $s$, must be the same for a vector space and its associated scalars. The reason is that $\bar{V}^{s}$ includes in its description a map, $\bar{S}^{s}\times \bar{V}^{s}\rightarrow \bar{V}^{s}$, as scalar vector multiplication, and a norm map $\bar{V}^{s}\rightarrow Re\bar{S}^{s}.$ Here $Re\bar{S}^{s}$ is the real part of $\bar{S}^{s}$ if $S=C.$

The fiber bundle, $\mathfrak{SV},$  is a principal fiber bundle.  This follows from the discussion in subsection \ref{RI}. It also is a consequence of the fact that there is a structure group, $W_{SV}$, for the fiber \cite{Drechsler}. This can be best seen by  separate examination of the principal  bundle for scalars, \begin{equation}\label{MFS}\mathfrak{S}=M,\bigcup_{s}\bar{S}^{s},q_{s},M. \end{equation} The structure group, $W_{S},$ for this bundle consists of the elements $W_{S,t}$ where $t$ is a number in the group, $GL(1,S).$ The group, $W_{S}$ acts freely and transitively \cite{Drechsler} on the fiber $\bigcup_{s}\bar{S}^{s}$ in that\begin{equation}\label{WSt}\begin{array}{l}W_{S,t}\bar{S}^{s}=\bar{S}^{st} =\bar{S}^{ts}\mbox{  and}\\\\ W_{S,u}(W_{S,t}\bar{S}^{s})=W_{S,u}\bar{S}^{ts}= \bar{S}^{uts}.\end{array}\end{equation}Here $GL(1,S)$ stands for $GL(1,C),$ the linear group of all complex numbers if $S=C.$  For real numbers with $S=R,$ the group is $GL(1,R\geq 0)$,  the linear group of all nonnegative real numbers. For $t=0,$ the structure, $\bar{S}^{0},$ is the empty structure.

 The  value preserving map, $W_{S,u},$ is related to $W_{S,s,t}$ of Eq. \ref{WSst} by \begin{equation}\label{WSu}W_{S,u}\bar{S}^{t} =W_{S,tu,t}\bar{S}^{t}=\bar{S}^{tu}. \end{equation}  A similar map, $W_{SV,u}$ is related to $W_{SV,s,t}$  in Eq. \ref{WSVst} by \begin{equation}\label{WSVu}W_{SV,u}(\bar{S}^{t}\times \bar{V}^{t}) =W_{SV,tu,t}(\bar{S}^{t}\times\bar{V}^{t})= \bar{S}^{tu}\times\bar{V}^{tu}. \end{equation}

 The group, $W_{SV}$, is the structure group for the fiber bundle $\mathfrak{SV}.$ The elements, $W_{SV,t}$ of $W_{SV}$ act freely and transitively on the fiber $\bigcup_{s}(\bar{S}^{s}\times\bar{V}^{s}))$.  This is  expressed by Eq. \ref{WSVu}  and \begin{equation} \label{WSVut}W_{SV,u}(W_{SV,t}(\bar{S}^{s} \times\bar{V}^{s}))=W_{SV,u}(\bar{S}^{ts}\times\bar{V}^{ts})=\bar{S}^{uts}\times\bar{V}^{uts}.\end{equation}

\section{Connections and sections}\label{CS}
The description of physical and geometric quantities as integrals or derivatives on $M$ requires the use of connections \cite{Mack}.  Connections, as maps between structures of the same type in  different  fibers, relate the contents of a fiber at one location to those of another at a different location. They describe the parallel transport of structures between different fibers.

Connections are necessary in the description of any nonlocal physical quantity. The reason is that  mathematical combinations, such as addition, multiplication, etc., of quantities in different fibers are not defined.  The quantities must be parallel transferred to one fiber before combinations are possible. In addition, connections  describe the effect of external fields on physical systems whose dynamics is described by motion along a path in $M$.  The state of a system at a given point is an element of a structure in a fiber at the point.

The usual fiber bundle framework for discussing physical fields in gauge theories has each fiber containing $\bar{S}\times\bar{V}.$ This would correspond here to having each fiber containing $\bar{S}^{t}\times \bar{V}^{t}$ for $t=1$ only. The fiber bundles described here are different in that each fiber contains the structure pairs, $\bar{S}^{t}\times\bar{V}^{t}$ for all $t$ in $GL(1,S)$.

Connections between fibers in the bundle, $\mathfrak{SV},$ relate  scalar vector structure pairs in one fiber to those in another.  The detailed description of connections given here is an extension of the usual description in gauge theories.  A description for scalars based on the fiber bundle, $\mathfrak{S},$ is followed by one for scalars and vectors based on the bundle, $\mathfrak{SV}.$

\subsection{Scalar structure valued sections}\label{SSVS}

Let $\Psi$ be a section  over the bundle $\mathfrak{S}.$ For each $x$ in $M$, $\Psi(x)$ is a scalar \emph{structure} in the fiber at $x.$ It is not a scalar or vector. $\Psi$ can also be considered as a scalar structure valued field over $M$.

To discuss properties of $\Psi$ such as smoothness, derivatives, etc., it is useful to associate a scalar function, $f$ with domain $M$ and range in the set of fiber level labels as elements of $GL(1,S).$  The function $f$ can also be considered as a $GL(1,S)$ valued scaling field over $M$. There is a one-one correlation between the set of functions, $f$ and sections $\Psi$ where \begin{equation}\label{PsixbS}\Psi(x)=\bar{S}^{f(x)}_{x} =\Psi_{f}(x).\end{equation}  The scalar structure, $\bar{S}^{f(x)}_{x},$ is at level $f(x)$ in the fiber at $x.$ The structure, $\bar{S}^{f(x)}_{x},$ is defined by Eq. \ref{BSs} with $s=f(x).$ This representation of $\Psi$ can be used to describe properties of $\Psi$ in terms of those of $f.$

\begin{figure}\begin{center}\rotatebox{270}{\resizebox{4.5cm}{4.5cm}
  {\includegraphics[7cm,8cm][18cm,19cm]{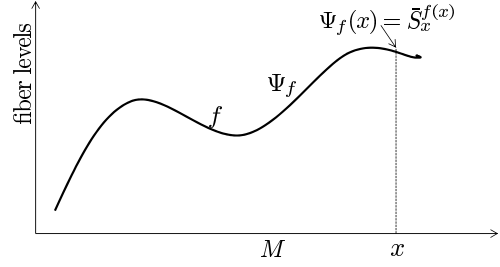}}}\caption{A schematic representation of both the level valued function or field and the associated  scalar structure field, $\Psi_{f}$. At each point, $x,$ of $M,$ $f(x)$ is a scaling level in the fiber at $x$.  $\Psi_{f}(x)$ is the scalar structure, $\bar{S}^{f(x)}_{x},$ at the level given by $f.$ This same picture holds for the scalar and vector section in that $\Psi_{SV,f}(x)=\bar{S}^{f(x)}_{x} \times\bar{V}^{f(x)}_{x}.$ }\label{FBA1}\end{center}\end{figure}

Figure \ref{FBA1} shows the relation between $\Psi$ and $f$ as curves in  a two dimensional space. The ordinate is labelled with  the values of $s$ or fiber levels in the bundle, and abscissa is labelled with points of $M$. Points in the space are both values, $f(x),$ of $f$ and scalar structures, $\bar{S}^{f(x)}_{x}.$

A good example in which to see the relations and parallel transports between these quantities is provided by the derivative of $\Psi.$ This is given by \begin{equation}\label{dPsi}d\Psi=\partial_{x}\Psi(x) dx +\partial_{f(x)}\Psi(x)df(x)=\partial_{\mu,x} \bar{S}^{f(x)}_{x}d^{\mu}x+\partial_{\mu,x,f} \bar{S}^{f(x)}_{x}d^{\mu}f(x).\end{equation} Summing over repeated indices is implied.

The partial derivatives are defined by  \begin{equation}\label{Pmux}\begin{array}{c}\partial_{\mu,x}\bar{S}^{f(x)}_{x}= \lim_{d^{\mu}x\rightarrow 0}\frac{\mbox{$\bar{S}^{f(x)}_{x+d^{\mu}x} -\bar{S}^{f(x)}_{x}$}}{\mbox{$d^{\mu}x$}} \\\\\partial_{\mu,x,f} \bar{S}^{f(x)}_{x}=\lim_{d^{\mu}f(x)\rightarrow 0}\frac{\mbox{$\bar{S}^{f(x) +d^{\mu}f(x)}_{x}- \bar{S}^{f(x)}_{x}$}}{\mbox{$d^{\mu}f(x)$}}.\end{array}\end{equation} These equations use the observation that $f(x+d^{\mu}x)=f(x)+d^{\mu}f(x).$

There are  problems with the definitions of the partial derivatives.  Scalar structures can be subtracted if and only if they are at the same location in $M$ and  at the same level in a fiber.  These problems can be fixed by use of connections to parallel transfer  $\bar{S}^{f(x)}_{x+d^{\mu}x}$ from $x+d^{\mu}x$ to $x$  and to  transfer  $\bar{S}^{f(x)+ d^{\mu}f(x)}_{x}$ to a structure whose components are expressed in terms of those of $\bar{S}^{f(x)}_{x}.$  The transfer from $x+d^{\mu}x$ to $x$ is defined from Eq. \ref{BSstw} by \begin{equation} \label{BSfxxd}\bar{S}^{f(x)}_{x+d^{\mu}x}\rightarrow w(x,x+d^{\mu}x)\bar{S}^{f(x)}_{x}=\bar{S}^{f(x)w}_{x,f(x)}. \end{equation}  Here $w=w(x,x+d^{\mu}x)$.

Replacing $\bar{S}^{f(x)}_{x+d^{\mu}x}$ by $\bar{S}^{f(x)w}_{x,f(x)}$ in the first of  Eqs. \ref{Pmux} and use of Eqs. \ref{BStpmu} and \ref{BSstw} gives for the covariant derivative, \begin{equation}\label{DmuxSw}
D_{\mu,x}\Psi(x)=\bar{S}^{f(x)(w-1)/d^{\mu}x}_{x,f(x)}.\end{equation}The connection coefficient, $w(x,x+\vec{dx}),$ is defined in terms of two vector fields as \begin{equation}\label{wxwf}
w(x,x+\vec{dx})=e^{(\vec{A}(x)+i\vec{B}(x))^\cdot\vec{dx}}\end{equation} The $\mu$ components are given by \begin{equation}\label{wxwfmu}w(x,x+d^{\mu}x)=e^{(A_{\mu}(x)+iB_{\mu}(x)d^{\mu}x)}. \end{equation}

Expansion of the exponential in $w$ to first order and carrying out the indicated operations in the superscript gives as the final result, \begin{equation}\label{DmuAB}D_{\mu,x}\Psi(x)=\bar{S}^{f(x)(A_{\mu}(x) +iB_{\mu}(x))}_{x,f(x)}=(A_{\mu}(x)+iB_{\mu}(x))\Psi(x).\end{equation}

The   effect of fiber level change  from $f(x)+d^{\mu}f(x)$ to $f(x)$ at $x$ on $\bar{S}^{f(x)+d^{\mu}f(x)}_{x}$   in the second of Eqs. \ref{Pmux} is given by \begin{equation}\label{ZBSfx}
\bar{S}^{f(x)+d^{\mu}f(x)}_{x}\rightarrow Z_{f(x),f(x+d^{\mu}f(x)}\bar{S}^{f(x)+d^{\mu}f(x)}_{x}= \bar{S}^{f(x)+d^{\mu}f(x)}_{f(x),x}.\end{equation} the map $Z_{f(x),f(x)+d^{\mu}f(x)}$ is defined by Eq. \ref{ZSst}.

The resulting covariant derivative is given by \begin{equation}\label{Dmuf}D_{\mu,x,f}\Psi(x)= \frac{\mbox{$\bar{S}^{f(x) +d^{\mu}f(x)}_{x,f(x)}- \bar{S}^{f(x)}_{x}$}}{\mbox{$d^{\mu}f(x)$}}.\end{equation} A Taylor expansion of $f((x)+d^{\mu}f(x)$, retention of first order terms, and use of Eqs. \ref{BStpmu} and \ref{BSstw}  gives, \begin{equation}\label{Dmfp}D_{\mu,x,f}\Psi(x)=\frac{\partial_{\mu,f}f(x)}{f(x)}\Psi(x).
\end{equation} Representation of the field $f$ as the exponential of a pair of scalar fields as in
\begin{equation}\label{fxegp}f(x)=e^{\gamma(x)+i\phi(x)}\end{equation} gives as a final result for the covariant derivative \begin{equation}\label{DmuGD}D_{\mu,x,f}\Psi(x)=(\Gamma_{\mu}(x)+i\Delta_{\mu} (x))\Psi(x).\end{equation}Here $\vec{\Gamma}(x)$ and $\vec{\Delta}(x)$ are the gradients of $\gamma(x)$ and $\delta(x).$  Note that the fields $\gamma$ and $\delta$, and their gradients, serve as connections between different levels in each fiber of the bundle, $\mathfrak{S}.$

These results apply to both real and complex scalar structures.  If $S=R$ then $\vec{B}$ and $\vec{\Delta}$ are $0$ everywhere on their domains. If $f(x)=s$, a constant, then $\gamma$ and $\delta$ are $0$ on their domains.

\subsection{Scalar and vector structure valued sections}\label{SVSVS}

The description of scalar structure valued sections or fields and their derivatives extends in a straight forward manner to fields that are scalar and vector structure valued. These fields are sections of the fiber bundle, $\mathfrak{SV}$, described in Eq. \ref{MFSV}.

Let $\Psi_{SV}=\Psi_{S}\times\Psi_{V}$ be a section on $\mathfrak{SV}$ whose value, in the fiber at $x$,  is given by \begin{equation}\label{PsiSVx}\Psi_{SV}(x)=\Psi_{S}(x)\times \Psi_{V}(x)=\bar{S}_{x}^{f(x)} \times\bar{V}_{x}^{f(x)}.\end{equation} The factor, $\Psi_{S},$ is identical to the section $\Psi$ on the bundle, $\mathfrak{S}.$ The definition of $\Psi_{SV}$ shows that the scaling or value parameter, $f(x)$, is the same for both scalar and vector structures.  This is  required by the fiber structure of $\mathfrak{SV}.$

The derivative of $\Psi_{SV}$ is given by \begin{equation}\label{pmuxPSV}\partial_{\mu,x}\Psi_{SV}= \partial_{\mu,x} (\Psi_{S}(x) \times\Psi_{V}(x))=\partial_{\mu,x}\Psi_{S}(x)\times\Psi_{V}(x)+\Psi_{S}(x)\times\partial_{\mu,x}\Psi_{V}(x) \end{equation} The same problems leading to the replacement of  partial derivatives for $\Psi_{s}$ with covariant derivatives apply to $\Psi_{V},$   and they are solved in the same way.  The result is given by\begin{equation}\label{DmuxPSV}D_{\mu,SV}\Psi_{SV}(x)= D_{\mu,S}\Psi_{S}(x)\times\Psi_{V}(x) +\Psi_{S}(x)\times D_{\mu,V}\Psi_{V}(x). \end{equation}Here $D_{\mu,S}\Psi_{S}(x)$ is  the sum of two component derivatives as in  \begin{equation}\label{DmuS}D_{\mu,S}\Psi_{S}(x)=(D_{\mu,x}+D_{\mu,x,f})\Psi_{S}(x).
\end{equation}The two component derivatives are given by Eqs. \ref{DmuAB} and \ref{DmuGD}.

The covariant derivative, $D_{\mu,V}\Psi_{V}(x)$   is obtained in the same way  as is that for $\Psi_{S}$. The result is a pair of covariant derivatives .  These are given by  Eqs. \ref{DmuAB} and \ref{DmuGD} after  replacing $S$ with $V$.    The result is \begin{equation}\label{DmuV}D_{\mu,V}\Psi_{V}(x)= (D_{\mu,x}+D_{\mu,x,f})\Psi_{V}(x).\end{equation}

At this point the physical relevance, if any, of the  four fields is not known.  The $\vec{\Gamma}$ and $\vec{\Delta}$ fields are integrable as they are gradients of scalar fields.  It is not known if the $\vec{A}$ and $\vec{B}$ fields are integrable or not.  As will be seen later in a discussion of gauge theory, the $\vec{B}$ field may be the electromagnetic field.  In this case it is not integrable.

\subsection{Scalar and vector  valued fields}\label{SVF}
So far fields that were also sections on the fiber bundles, $\mathfrak{S}$ and $\mathfrak{SV}$, have been described. One can also describe scalar and vector valued fields that take values in the scalar and vector structures in the fibers of the bundles. These are the more usual types of fields described in detail in  gauge theories and in physics in general.  They are also not sections over the fiber bundles.

Let $\psi_{S}$ and $\psi_{V}$ denote scalar and vector valued fields.  For each $x$ in $M$, $\psi(x)$  and $\psi_{V}(x)$ are respective scalar and vector values in $\bar{S}^{f(x)}_{x}$ and $\bar{V}^{f(x)}_{x}.$  Note that the field values are values in their respective structures.  They are not elements of the base sets in the structures.

The fiber bundle framework enables  a description of fields that are more general than those described so far in gauge theories.  The is due to the presence of the function $f$.  The consequence of this is that the covariant derivatives appearing in Lagrangians and equations of motion include additional fields arising from the connections between fiber levels.   A description for scalar fields in $\mathfrak{S}$ will precede a description for scalar and vector fields in $\mathfrak{SV}.$

\subsubsection{Scalar fields}\label{SF}
The usual derivative, \begin{equation}\label{parmuxp}\partial_{\mu,x}\psi_{S}(x)=\frac{\psi_{S}(x+d^{\mu}x) -\psi_{S}(x)}{d^{\mu}x},\end{equation} is not defined because $\psi_{S}(x+d^{\mu}x)$ is a scalar in $\bar{S}^{f(x+d^{\mu}x)}_{x+d^{\mu}x}$ and $\psi(x)$ is a scalar in $\bar{S}^{f(x)}_{x}.$  This can be remedied by parallel transforming $\psi_{S}(x+d^{\mu}x)$ to a value in $\bar{S}^{f(x+d^{|mu}x)w(x,x+d^{\mu}x)}_{x,f(x)}.$ The reason this works is that the elements of this structure are scaled elements of $\bar{S}^{f(x)}_{x}.$

The resulting  covariant derivative of $\psi_{S}$  defined by \begin{equation}\label{DpsSfr}D_{\mu,x}\psi_{S}= \frac{\frac{t}{s}(\psi_{S} (x+d^{\mu}x))-\psi_{S}(x)}{d^{\mu}x}.\end{equation}Here $t=w(x,x+d^{\mu}x)f(x+d^{\mu}x)$ and $s=f(x).$ The connection, $w(x,x+d^{\mu}x),$ is given by Eq. \ref{wxwfmu}. Expansion of the exponential, combined with the Taylor expansion of $f(x+d^{\mu}x),$ and retention of terms up to and including first order in $d^{\mu}x$ gives\begin{equation}\label{DmuxpsiS}D_{\mu,x}\psi_{S}= (\partial_{\mu,x}+A_{\mu}(x)+iB_{\mu}(x)+ \frac{\partial_{\mu,x}f(x)}{f(x)})\psi_{S}.\end{equation} Expression of $f(x)$ as the exponential of a complex scalar field, as in Eq. \ref{fxegp}, gives the final expression for the covariant derivative as \begin{equation}\label{DmuxABGD}D_{\mu,x}\psi_{S}= (\partial_{\mu,x}+A_{\mu}(x)+iB_{\mu}(x)+ \Gamma_{\mu}(x)+i\Delta_{\mu}(x))\psi_{S}.\end{equation} Here $\vec{\Gamma}$ and $\vec{\Delta}$ are  gradients of the scalar fields, $\gamma$ and $\delta.$ If needed, coupling constants for the four fields can be added.

This result differs from covariant derivative for the scalar structure valued field, $\Psi_{S}$ by the presence of the partial derivative, $\partial_{\mu,x}\psi_{S}.$  This is to be expected because scalar  field values can vary even if they are referenced to the same scalar structure. If $\psi$ and $\psi'$ are two different scalar fields, then $\psi(x)\neq \psi'(x)$ is possible even though both $\psi(x)$ and $\psi'(x)$ are number values in $\bar{S}^{f(x)}_{x}.$

 Note that, if $y$ is different from $x,$ the expression for the covariant derivative, given by Eq. \ref{DmuxABGD} with $y$ replacing $x,$ is based on number values in $\bar{S}^{f(y)}_{y}$.  In general this structure is at a different  level in the fiber than is the expression at $x$ which is in $\bar{S}^{f(x)}_{x}.$  The level location of $D_{\mu,x}\psi_{S}$ is independent of $x$ if and only if $f$ is a constant everywhere.  In this case $\psi_{S}$ can be called a horizontal field as all of its values are in structures at the same level in the fibers of $\mathfrak{S}.$

One should keep in mind that, for each $x,$ Eq. \ref{DmuxABGD} is an expression involving number values in $\bar{S}^{f(x)}_{x}.$ This can be made explicit by adding the subscript, $f(x),$ to each term in the equation. Implied operations, such as multiplication and addition, also have subscripts as in $\times_{f(x)}$ and $+_{f(x)}.$

\subsubsection{Vector fields}\label{SAVF}
Vector valued fields differ from scalar valued fields in that they take values in the structures $\bar{V}^{f(x)}_{x}$  for different $x$ in $M$. As is seen by the fiber structure in the bundle, $\mathfrak{SV}$ in Eq. \ref{MFSV}, $\bar{S}^{f(x)}_{x}$ is the scalar structure associated with $\bar{V}^{f(x)}_{x}.$

The derivative, $\partial_{\mu,x}\psi_{V}$ of a vector valued field $\psi_{V}$ has the same problems as does the derivative of a scalar valued field, $\psi_{S}$. The problems are remedied in the same way as was done for the scalar field.  The resulting expression for the covariant derivative  is given by \begin{equation} \label{DmuxpsiV}D_{\mu,x}\psi_{V}=(\partial_{\mu,x}+A_{\mu}(x)+\Gamma_{\mu}(x)+i(B_{\mu}(x)+ \Delta_{\mu}(x)) \psi_{V}.\end{equation} This is the same expression as that for the scalar field in Eq. \ref{DmuxpsiS}. As was the case for the scalar field, coupling constants can be added to the four fields.

The covariant derivative, $D_{\mu,x}\psi_{V},$ accounts for the scalar degrees of freedom associated with the changes in fiber level and location on $M$.  It does not account for changes in bases for the vector spaces.  These are included in gauge theories.

\section{Gauge theories}\label{GT}

As is well known \cite{Yang,Utiyama,Cheng}, covariant derivatives of vector fields replace the usual derivatives in Lagrangians and equations of motion. The usual framework for these theories corresponds to the special case here where $f(x)=1$ for all $x$ in $M$ and $\vec{A}$ and $\vec{B}$ are $0$ everywhere.  If $\bar{V}$ is $N$ dimensional then $U(N)=U(1)\times SU(N)$ is the gauge group. The manifold $M$ is taken to be space time.

Descriptions of fields in gauge theories assign separate vector spaces to each point of $M$. (This is shown explicitly in \cite{Montvay}.)  Vector fields $\psi_{V}$ take values in these spaces with $\psi_{V}(x)$ a vector in the space,  $\bar{V}_{x}.$ As a result the usual derivatives, which require comparison of vector values in neighboring spaces, are undefined.  This problem is solved by replacing usual derivatives with covariant derivatives. These relate fields values at neighboring points by use of connections or gauge group elements  as parallel transfers \cite{Mack}. For the gauge group $U(N)$, the  covariant derivative is given by \begin{equation}\label{DpsiV} D_{\mu,x}\psi_{V} =(\partial_{\mu,x} +ig_{1}E_{\mu}(x)+\frac{ig}{2}\alpha^{j}_{\mu}(x)\tau_{j})\psi_{V}.\end{equation} Here $\vec{E}$ is the electromagnetic field and $(ig/2)\alpha^{j}_{\mu}(x)\tau_{j}$ is the Lie algebra, $su(N)$ representation of an element of $SU(N).$  The $j$ sum is over all $N^{2}-1$ generators, $\tau_{j},$ of the algebra. The $\alpha^{j}_{\mu}(x)$ are real numbers.  Coupling constants, $g_{1}$ and $g,$ have been added.\footnote{These concerns were the origin of the work described in the previous sections and in other work by the author.}

Invariance of terms in Lagrangians under local $U(1)$ gauge transformations, is achieved if the dependence of $\vec{E}$ on local gauge transformations, $U(x),$ is given by\begin{equation}\label{vBp}E_{\mu}'(x) =U^{\dag}(x) E_{\mu}(x)U(x) +\frac{1}{g}U^{\dag}(x)\partial_{\mu,x}U(x). \end{equation}  The $\vec{E}'$ field is  massless.

This description applies to Abelian gauge theories.  For nonabelian theories, with $N=2$, invariance of Lagrangian terms under local $SU(2)$ gauge transformations, $U(x),$  means that the vector fields $\vec{\alpha}^{j}$ transform according to \cite{Cheng} \begin{equation}\label{vecal}\vec{\alpha}'_{\mu}\cdot \mathbf{\tau}=U(x)\vec{\alpha}_{\mu}\cdot\mathbf{\tau} U^{\dag}(x)+\frac{2i}{g}(\partial_{\mu,x}U(x)) U(x)^{\dag}.\end{equation} Here $\mathbf{\tau}$ are the three Pauli operators and the $\vec{\alpha}^{j}$ for $j=1,2,3$ are three vector boson fields.

This brief summary of the setup for gauge theory, translated into the fiber bundle setup described in the earlier sections,  is for the special case of a constant scaling field $f$  with $f=1$ everywhere, and $\vec{A}$ and $\vec{B}=0.$   The corresponding  fiber bundle basis  is \begin{equation}\label{MFCV1}\mathfrak{CV}_{1}=M,(\bar{C}\times\bar{V}),p,M\end{equation} with an associated principal frame bundle.

 Expansion of the bundle setup for gauge theories  to $\mathfrak{SV}$, Eq. \ref{MFSV}, results in an expansion of the gauge theory covariant derivatives to include terms for the scalar degrees of freedom, as in Eq. \ref{DmuxABGD}. Here  $S=C$  and $f$  is a complex valued field.   An associated principal frame bundle also must be included.

The bundle levels at which vector fields take values are determined by $f$. A vector field, $\psi$ has values $\psi(x)$ in $\bar{V}^{f(x)}_{x}.$ The scalars for $\bar{V}^{f(x)}_{x}$ are those in $\bar{C}^{f(x)}_{x}.$  The covariant derivative at $x$ is at level $f(x)$. It is given by an expansion of  Eq. \ref{DmuxpsiV} to \begin{equation} \label{Dmuxpal}D_{\mu,x}\psi=(\partial_{\mu,x}+g_{a}A_{\mu}(x)+ig_{b}B_{\mu}(x)+g_{g}\Gamma_{\mu}(x)+ +ig_{d}\Delta_{\mu}(x)+\frac{ig}{2}\alpha^{j}_{\mu}(x)\tau_{j})\psi_{V}.\end{equation}Separate coupling constants have been added for the four fields.

The associated gauge group for the covariant derivative in Eq. \ref{Dmuxpal} is $GL(1,C)\times GL(1,C)\times SU(N).$ The connections, $\exp ((\vec{A}+i\vec{B})\cdot \vec{dx})$ and $\exp ((\vec{\Gamma}+i\vec{\Delta})\cdot \vec{x})$ are elements of the two $GL(1,C).$

It is assumed here that  $\vec{B}$ is the electromagnetic field, $\vec{E},$  of Eq. \ref{DpsiV}. The field, $\vec{\Delta}$ cannot  assume this role because it is integrable. This role exclusion is a consequence of the Aharonov Bohm effect \cite{AhBo}.

The assumption that $\vec{B}$ is the electromagnetic field is based on the assumption that the $U(1)$ component of  the usual $U(N)$ gauge group is included in  the  scalar  gauge group associated with the $\vec{A}$ and $\vec{B}$ fields. This assumption can be excluded by expansion of the overall gauge group to $GL(1,c)\times GL91,C)\times U(N).$ This would add another vector field, $\vec{E}$  term to the  expression for the covariant derivative.

The separation of $\vec{E}$ from $\vec{B}$ does require that the coupling constants not be the same for the two fields.  If they are the same, then, as far as gauge theory is concerned, they represent an arbitrary division of one field into the sum to two fields.

The expansion of the gauge group to $GL(1,C)\times GL(1,C)\times U(N)$  would have the consequence  that the relevance, if any, of all four fields, $\vec{A},\vec{B},\vec{\Gamma},\vec{\Delta}$  to physics would  not be known.  Even with $\vec{B}$ the electromagnetic field, this problem remains for the other three fields.\footnote{A very speculative suggestion \cite{BenFB} is that $\vec{A}$ is the gradient of the Higgs field.} This, and other aspects, are topics for future work.

  \section{Discussion}

  This are many  avenues open for expansion of  this work.  One is the replacement of $M$ as a flat space to be a Riemannian or pseudo Riemannian manifold. In this case the power of the fiber bundle approach becomes more evident as the bundles are product bundles on local regions of $M$ but not on all of $M$. This would be the setup in which to include general relativity.

   The construction of the bundle $\mathfrak{SV}$ from the individual scaled bundles $\mathcal{SV}_{s}$ is an example of how one can expand fiber bundles so that each fiber contains more and more types of mathematical systems. There are many ways the fibers can be expanded. One can include structures for vector spaces based on both real and on complex scalars.  Other mathematical systems that use scalars in their description,  such as algebras and groups, can be included.

   For each of the expansion examples discussed here, and for many others,  the fiber at each point, $x,$ of $M$, contains structures for many different types of mathematical systems.  These can all be considered as structures localized at $x$.\footnote{These fibers give a more precise description of the universes of mathematics at each point of $M$ described in earlier work\cite{BenIntech,BenNova}.}

   Another interesting expansion is to include in each  fiber a representation or model of $M$.  This can be done by defining at each point $x$ of $M$ a chart, $\phi_{x},$ that maps $M$ onto  a structure, $\bar{R}^{n}_{x},$ in the fiber at $x$. Charts are  open set preserving, one one maps of $M$ onto $\bar{R}^{n}.$  Since $M$ is flat, an atlas of charts over open subsets of $M$ is not needed.  Single charts on all of $M$ are sufficient.\footnote{Some aspects of the family of charts, $\phi_{x},$ at all points of $M$ have been discussed elsewhere \cite{BenSpie4}.}

   An interesting use of this expansion is that it  would enable the lifting of global descriptions of fields over $M$ to local descriptions as fields over $\bar{R}^{n}_{x}$ in the fiber at $x.$ Covariant derivatives and other aspects of  gauge field theories could then be described locally. This  is clearly an area in which much can be done.

   A major open problem is the relation between the base sets of structures and physical sets of symbol strings.  As emphasized  earlier, elements of base sets are meaningless when considered by themselves.  They acquire meaning only in the context of a structure for which relevant axioms are true. This problem has already showed up in the labeling of base set elements as $sa$ in structures $\bar{S}^{s}$ as in subsection \ref{SS}.  If $a$ and $s$ are  variable labels for elements of the base set, $S$, then $sa$ implies multiplication of these elements.  But multiplication in $\bar{S}^{s}$ is defined on number values, not on the base set elements.

 A possibly useful avenue of approach to this problem is to note that it is  similar to the description of syntactic structures in mathematical logic \cite{Barwise}  as meaningless sets of symbol strings and their relation to semantic or meaningful structures.  Some aspects of this have been noted in subsection \ref{DC}.

 It was noted in the introduction that the concepts of number and number value can be identified as long as one considers single structures for the different types of numbers.  This is in  accord with the Platonic view of mathematical elements in that each one has an ideal existence with unique properties.\footnote{This cannot be true for  all numbers in $R$ or $C$ but only for a dense subset (rationals).} for example the numbers, $\pi$ or $4.72 +i\pi$, as elements of the base sets,  have unique properties  independent of their belonging to a base set in a structure.

 This is quite like the classical mechanical or Newtonian view of physical systems.  In this view a particle has a momentum and a position that exists  independent of any measurement of  the particle's position or momentum. Measurements consist of determining what the values are of the position and momentum.  A measurement of the value of these  quantities corresponds to selecting a pair of numbers in $R$ whose values are  intrinsic  to  the particle at a given point in time.

 The position taken here is rather different. Here the elements of the base sets $R$ and $C$ can be regarded as numbers.  However, considered by themselves, there are no values associated with the elements of the base sets.  Each element can have any value or possibly none at all. The elements of $R$ or $C$ acquire value or meaning only as base sets contained in  structures  that satisfy relevant sets of axioms.

 This is more in accord with the quantum mechanical view of properties of physical systems. A particle, described by a wave function, has no specific position or momentum.  Any values are possible, subject to the position momentum uncertainty principle.  A single measurement on a system gives, as outcomes, values of these quantities. Repeated measurements on different identically prepared systems, give different outcome values. The probabilities of different values are properties obtained from the wave function.

 The correspondence with base sets $R$ and $C$ is that, like the wave function particle description,  any value is possible for the elements of the base sets when considered by themselves.  A single  measurement of particle position or momentum, yields a value as an outcome.  This is like measuring the value of an element of the base set where the outcome is a value for the  measured element.  This is equivalent to choosing the value structure in which the base set element measured has "measured" value.

 Further exploration of the correspondence between particle mechanics and base sets in number structures is left to the future.  Here the main point is that, by themselves, the elements of the base sets have no specific values.  They acquire values only when contained in a structure. The values of the elements are different in different structures. However, the totality of values is the same for all structures.

  The work described in this paper is part of a general problem of constructing a coherent theory of mathematics and physics together.  It is hoped that this work will lead to a better understanding of how to approach this very interesting problem.

  \section*{Acknowledgement}
  This material is based upon work supported by the U.S. Department of Energy, Office of Science, Office
  of Nuclear Physics, under contract number DE-AC02-06CH11357.

\end{document}